\documentclass[11pt]{article}
\usepackage[margin=1in]{geometry}
\usepackage{cite}
\usepackage{amsmath,amssymb,amsfonts}
\usepackage{outlines}
\usepackage{caption}
\usepackage{subcaption}
\usepackage{authblk}
\usepackage{algorithmic}
\usepackage{graphicx}
\usepackage[T1]{fontenc}
\usepackage{booktabs}
\usepackage{fontawesome5}
\usepackage{float}
\usepackage{array}
\usepackage{multirow}
\usepackage{textcomp}
\usepackage{xcolor}

\definecolor{lightgray}{rgb}{0.95, 0.95, 0.95}
\definecolor{darkgray}{rgb}{0.4, 0.4, 0.4}
\definecolor{editorGray}{rgb}{0.95, 0.95, 0.95}
\definecolor{editorOcher}{rgb}{1, 0.5, 0} %
\definecolor{editorGreen}{rgb}{0, 0.5, 0} %
\definecolor{orange}{rgb}{1,0.45,0.13}
\definecolor{olive}{rgb}{0.17,0.59,0.20}
\definecolor{brown}{rgb}{0.69,0.31,0.31}
\definecolor{purple}{rgb}{0.38,0.18,0.81}
\definecolor{lightblue}{rgb}{0.1,0.57,0.7}
\definecolor{lightred}{rgb}{1,0.4,0.5}
\usepackage{upquote}
\usepackage{listings}
\usepackage{tabularx}
\usepackage{color}
\definecolor{lightgray}{rgb}{.9,.9,.9}
\definecolor{darkgray}{rgb}{.4,.4,.4}
\definecolor{purple}{rgb}{0.65, 0.12, 0.82}

\lstdefinelanguage{JS}{
  keywords={typeof, new, true, false, catch, function, return, null, catch, switch, var, if, in, while, do, else, case, break, maliciousCode},
  keywordstyle=\color{blue}\bfseries,
  ndkeywords={class, export, boolean, throw, implements, import, this},
  ndkeywordstyle=\color{darkgray}\bfseries,
  identifierstyle=\color{black},
  sensitive=false,
  comment=[l]{//},
  morecomment=[s]{/*}{*/},
  commentstyle=\color{purple}\ttfamily,
  stringstyle=\color{red}\ttfamily,
  morestring=[b]',
  morestring=[b]"
}

\lstdefinelanguage{CSS}{
  keywords={color,background-image:,margin,padding,font,weight,display,position,top,left,right,bottom,list,style,border,size,white,space,min,width, transition:, transform:, transition-property, transition-duration, transition-timing-function},
  sensitive=true,
  morecomment=[l]{//},
  morecomment=[s]{/*}{*/},
  morestring=[b]',
  morestring=[b]",
  alsoletter={:},
  alsodigit={-}
}

\lstdefinelanguage{JavaScript}{
  morekeywords={typeof, new, true, false, catch, function, return, null, catch, switch, var, if, in, while, do, else, case, break},
  morecomment=[s]{/*}{*/},
  morecomment=[l]//,
  morestring=[b]",
  morestring=[b]'
}

\lstdefinelanguage{HTML5}{
  language=html,
  sensitive=true,
  alsoletter={<>=-},
  morecomment=[s]{<!-}{-->},
  tag=[s],
  otherkeywords={
  >,
	<!DOCTYPE,
  </html, <html, <head, <title, </title, <style, </style, <link, </head, <meta, />,
	</body, <body,
	</div, <div, </div>,
	</p, <p, </p>,
	</script, <script,
  <canvas, /canvas>, <svg, <rect, <animateTransform, </rect>, </svg>, <video, <source, <iframe, </iframe>, </video>, <image, </image>, <header, </header, <article, </article
  },
  ndkeywords={
  =,
  charset=, src=, id=, width=, height=, style=, type=, rel=, href=,
  fill=, attributeName=, begin=, dur=, from=, to=, poster=, controls=, x=, y=, repeatCount=, xlink:href=,
  margin:, padding:, background-image:, border:, top:, left:, position:, width:, height:, margin-top:, margin-bottom:, font-size:, line-height:,
  transform:, -moz-transform:, -webkit-transform:,
  animation:, -webkit-animation:,
  transition:,  transition-duration:, transition-property:, transition-timing-function:,
  }
}

\lstdefinestyle{htmlcssjs} {%
  basicstyle={\footnotesize\ttfamily},
  frame=b,
  xleftmargin={0.75cm},
  numbers=left,
  stepnumber=1,
  firstnumber=1,
  numberfirstline=true,
  identifierstyle=\color{black},
  keywordstyle=\color{blue}\bfseries,
  ndkeywordstyle=\color{editorGreen}\bfseries,
  stringstyle=\color{editorOcher}\ttfamily,
  commentstyle=\color{brown}\ttfamily,
  language=HTML5,
  alsolanguage=JavaScript,
  alsodigit={.:;},
  tabsize=2,
  showtabs=false,
  showspaces=false,
  showstringspaces=false,
  extendedchars=true,
  breaklines=true,
  literate=%
  {Ö}{{\"O}}1
  {Ä}{{\"A}}1
  {Ü}{{\"U}}1
  {ß}{{\ss}}1
  {ü}{{\"u}}1
  {ä}{{\"a}}1
  {ö}{{\"o}}1
}
\lstdefinestyle{py} {%
language=python,
literate=%
*{0}{{{\color{lightred}0}}}1
{1}{{{\color{lightred}1}}}1
{2}{{{\color{lightred}2}}}1
{3}{{{\color{lightred}3}}}1
{4}{{{\color{lightred}4}}}1
{5}{{{\color{lightred}5}}}1
{6}{{{\color{lightred}6}}}1
{7}{{{\color{lightred}7}}}1
{8}{{{\color{lightred}8}}}1
{9}{{{\color{lightred}9}}}1,
basicstyle=\footnotesize\ttfamily, %
numbers=left,               %
numbersep=5pt,              %
tabsize=4,                  %
extendedchars=true,         %
breaklines=true,            %
keywordstyle=\color{blue}\bfseries,
frame=b,
commentstyle=\color{brown}\itshape,
stringstyle=\color{editorOcher}\ttfamily, %
showspaces=false,           %
showtabs=false,             %
xleftmargin=17pt,
framexleftmargin=17pt,
framexrightmargin=5pt,
framexbottommargin=4pt,
showstringspaces=false,      %
}%

\renewcommand*{\Affilfont}{\normalsize\normalfont}
\newsavebox\affbox

\usepackage{hyperref}
\usepackage{cleveref}
\title{Exposing and Addressing Security Vulnerabilities in Browser Text Input Fields}
\begin{document}

\newcommand*\samethanks[1][\value{footnote}]{\footnotemark[#1]}
\author{Asmit Nayak\thanks{Equal Contribution}}
\author{Rishabh Khandelwal\samethanks}
\author{Kassem Fawaz}
\affil[]{%
  \savebox\affbox{\Affilfont Department of Chemical Engineering, University of AAAAA BBBBBB, CCCCC road,}%
  \parbox[t]{\wd\affbox}{\protect\centering} University of Wisconsin -- Madison} 
\date{} 

\maketitle

\begin{abstract}
In this work, we perform a comprehensive analysis of the security of text input fields in web browsers. We find that browsers' coarse-grained permission model violates two security design principles: least privilege and complete mediation. We further uncover two vulnerabilities in input fields, including the alarming discovery of passwords in plaintext within the HTML source code of the web page. To demonstrate the real-world impact of these vulnerabilities, we design a proof-of-concept extension, leveraging techniques from static and dynamic code injection attacks to bypass the web store review process. Our measurements and case studies reveal that these vulnerabilities are prevalent across various websites, with sensitive user information, such as passwords, exposed in the HTML source code of even high-traffic sites like Google and Cloudflare. We find that a significant percentage (12.5\%) of extensions possess the necessary permissions to exploit these vulnerabilities and identify 190 extensions that directly access password fields. Finally, we propose two countermeasures to address these risks: a bolt-on JavaScript package for immediate adoption by website developers allowing them to protect sensitive input fields, and a browser-level solution that alerts users when an extension accesses sensitive input fields. Our research highlights the urgent need for improved security measures to protect sensitive user information online.
\end{abstract}

\section{Introduction}
Browser extensions are small applications that enhance the capabilities of web browsers and improve the user experience. They can add new features, modify web page content and automate tasks. Canonical examples of extensions include password managers, productivity tools such as Pocket, and ad blockers that modify webpages through well-defined APIs to prevent advertisements. The extensions achieve this by accessing the contents of the webpages and manipulating the contents of the webpages by executing JavaScript code.  
This access to web page content also allows the extension to retrieve private content, such as emails or banking details. 

Prior work has shown that it is possible to exploit this access to read sensitive user data such as emails~\cite{1_Liu2012ChromeET, 26_Varshney_detecting_2017}, passwords~\cite{8_bauer_analyzing_2014, 26_Varshney_detecting_2017}, and even perform phishing attacks~\cite{8_bauer_analyzing_2014, 4_perrotta_botnet_2018, 26_Varshney_detecting_2017}. These attacks can either use: a) \textit{Static Code Injection} where the attackers add the malicious code in the extension; or b) \textit{Dynamic Code Injection} where the code is loaded dynamically from a remote server and executed at run time. Static code injection is impractical as they can be detected by static code analysis~\cite{13_wang_combined_2018, 11_zhao_automatic_2015,12_dekoven_malicious_nodate,5_chen_mystique_2018}. Dynamic code injection bypasses the static security checks as the code is injected at run time, and thus, is harder to detect~\cite{3_kapravelos_hulk_nodate,21_Toreini_domtegrity_2019}. To address this vulnerability posed by dynamic code injection, Google introduced new regulations that disallowed the execution of remotely injected code. 

However, as we show in \Cref{sec:attack-design}, bypassing the protective measures and extracting sensitive information using the extension is possible. The attack is feasible because the interaction between the extensions and the web pages has not changed. The extensions can still access entire contents of the web pages, including text input fields where users may enter sensitive information such as passwords, Social Security Numbers (SSN), and Credit Card information.

In this work, we contribute to understanding the security of text input fields after Google's new regulations by performing an experimental analysis of the interactions between browser extensions and web pages. We focus on text input fields because input fields are often used for sensitive data such as usernames, passwords, credit card numbers, and SSNs. Given the sensitive nature of the data, ensuring that malicious actors cannot access input fields is of utmost importance. If an attacker can access or manipulate the data in these fields, they can potentially steal private user information, impersonate the user, or commit financial fraud. Exposure of this data could potentially also be harvested by automated scripts or bots that scan webpages for such vulnerabilities.

To conduct a systematic experimental analysis after introducing new regulations,  we first analyze the interactions between web extensions and web pages. We focus on the browser's permission model to isolate extensions (\Cref{sec:security-analysis}). We find that due to browsers' coarse-grained permission model, a lack of security boundary exists between the extension and the web page. This lack of boundary allows the extension to freely interact with and manipulate the HTML elements, including HTML input elements. Concretely, we have discovered that the extension permission model violates two security principles: 1) least privilege and 2) complete mediation. This allows a malicious extension to access sensitive user information without users' knowledge. 

To demonstrate the concrete harm posed to users, we first identify two vulnerabilities with input fields, including a novel vulnerability where sensitive information like passwords is visible in plain text in the HTML source code. Next, we explore how an attacker can exploit these vulnerabilities and gain access to private user information by designing a proof-of-concept extension(\Cref{sec:attack-design}). During the development of the extension, we introduce a novel attack to extract information from input fields. We leverage static and dynamic code injection techniques to create a hybrid attack that bypasses the webstore review process. 

Next, we conduct measurements (\Cref{sec:measurements}) and case studies (\Cref{sec:case_studies}) to understand the prevalence of these vulnerabilities.
We observe that both vulnerabilities exist within the input fields of the login forms across the websites. Specifically, we can extract usernames and passwords from the login pages of 100\% of the websites. Alarmingly, we find that on 15\% of the websites, passwords are present in plain text in the HTML source code. This implies that any entity, including extensions and third-party JavaScript, that can view the page source can extract users' passwords. This vulnerability is present in popular websites with high traffic, such as Google, Cloudflare, and Doordash. We further analyze the permissions requested by the extensions and find that 12.5\% of extensions have the necessary permissions to exploit the vulnerabilities discovered in this work. We also performed dynamic analysis by instrumenting the extensions and identifying 190 extensions directly accessing password fields. 

\noindent
\textbf{Contributions.} Our key contributions are summarized below:
\begin{itemize}
    \item We perform an experimental security analysis of the security of text input fields and find that the permission model violates the principle of least privilege and complete mediation. We also find two novel vulnerabilities in input fields, including one where passwords are present in the HTML source of the webpage. 
    \item We design proof-of-concept attacks to exploit these vulnerabilities and show the practicality of the attack by submitting the extension to the web stores.
    \item We perform measurements and case studies to show the prevalence of these vulnerabilities in the wild. We find the vulnerabilities across all websites indicating that websites fail to protect users' sensitive information.
\end{itemize}

Finally, we propose two countermeasures to mitigate the security risks from the observed vulnerabilities. In our \textit{Bolt-on} solution, we provide a JavaScript package that website developers can adopt today to mitigate the attacks (\Cref{subsec:bolt-on}). The package introduces a new input type \textit{SecureInput} that uses WeakMaps\footnote{\url{https://developer.mozilla.org/en-US/docs/Web/JavaScript/Reference/Global_Objects/WeakMap}} to store sensitive values in private variables. We also propose a more fundamental browser-level solution (\Cref{subsec:built-in}) by instrumenting chromium to alert users when an extension accesses sensitive input fields. 

\noindent
\textbf{Highlighting Severity of Vulnerability.} To highlight the severity of the vulnerability, we provide an example. Consider an adversary to steal login credentials from the users. Assume that the adversary can acquire a popular browser extension (as shown possible in~\cite{3_kapravelos_hulk_nodate, 22_Ursell_desktop_2019, redditAdwareMalware, arsTechnicaAdwareMalware}) - \textit{Honey\footnote{\url{https://honey.com/}}} with over 17M users. We note here that popular websites such as \url{google.com} and \url{cloudflare.com} also have the vulnerability where the passwords are visible in plain text in HTML source. As the extension has permission to run on all websites, the adversary can inspect the source code of the Google login page and extract the login credentials for millions of users.

\section{Background}
\label{sec:background}


We provide a brief background about the HTML fundamentals, including HTML input elements, HTML tree structure, accessibility of elements through JavaScript, and iframes. We also discuss the permission models for a browser extension, as well as the extension architecture.

\subsection{HTML Fundamentals}
\label{subsec:html-fundamentals}
\noindent
\textbf{HTML Input Elements:}
Input fields, marked by the HTML tag \texttt{<input>}, serve as the most basic avenue for users to input data into a webpage. These elements constitute a critical component of web forms\footnote{\url{https://developer.mozilla.org/en-US/docs/Learn/Forms}}. The rendering of the input field on the screen is dependent on the type of input field. For instance, the checkbox’ type establishes a checkbox, the text’ type generates a regular textbox, and the `password’ type, generally used for sensitive content, conceals the text shown on the textbox. We note that ensuring that input fields cannot be accessed by malicious actors is crucial, as exposure of sensitive data can be harvested by automated scripts or bots.


\smallskip

\noindent
\textbf{DOM Tree:}
While rendering a webpage, the browser constructs a Document Object Model (DOM) of the page. This DOM, composed of nodes and objects, replicates the webpage as a tree structure, known as the DOM Tree. The tree's root initiates with the \texttt{<html>} element.

The DOM API allows JavaScript (JS) to access any component of the DOM tree and adjust its attributes or content. This involves selecting the HTML element, which can be accomplished in several ways, such as by \textit{ID}, \textit{tag name}, or \textit{xpaths}. Moreover, the DOM API also enables JS to insert new elements into the DOM tree or discard existing ones.
%

\smallskip
\noindent
\textbf{Eval Statements:}
JavaScript allows the execution of strings as JavaScript code using the \texttt{eval()} function. While \texttt{eval} can be legitimately used to generate code based on specific conditions dynamically, its use is generally viewed as a security risk due to its potent nature. Richards et al.~\cite{27_Richards_eval_2011} performed a large-scale study on the use \texttt{eval} statements in web applications.

Extensions have been known to use \texttt{eval} statements to inject code into webpages dynamically. Kapravelos et al.~\cite{3_kapravelos_hulk_nodate} found more than 400 Chrome extensions using \texttt{eval} statements with inputs exceeding 128 characters in length. Similarly, Wang et al.~\cite{20_Wang_firefox_2012} discovered 145 extensions on the Firefox add-on store that contain the \texttt{eval} statement.

\smallskip
\noindent
\textbf{Inline Frames (iframe):} An inline frame, or iframe, is a special type of HTML element that can load another HTML page. Each iframe runs on its own context and cannot directly interact with the parent object or other iframes. Iframes are majorly used to load content from another page, this functionality can lead to security risks as misuse of iframes leads to a security vulnerability for users.



\subsection{Browser Extensions}
\label{subsec:architecture-of-browser-extensions}

Browser extensions provide a customizable user experience, adjusting web content to align with user preferences. Their functionalities range from aesthetic alterations like background color changes to ad removal and the automatic deletion of tracking cookies.  To accomplish these tasks, extensions require access to browser resources, APIs, and webpage content.  

\smallskip
\noindent
\textbf{Permission Models:} Browser extensions request permissions for the resources they require  for their functionality. requested via the manifest.json file, can be of two types: Host permissions and API permissions. Host permissions enable extensions to inform the browser about the websites they need to access, allowing extensions to access content from these specified sites. An extension can request access to a specific set of websites, or request access to all urls. API permissions, on the other hand, provide extensions with the capability to interact with specific WebExtension APIs, such as \texttt{browser.storage} or \texttt{browser.cookies}. 

\smallskip
\noindent
\textbf{Content Scripts and Background Pages:} Extensions are composed of two main components: content scripts and background pages (or service workers). Content scripts are static JavaScript files that are automatically loaded with a webpage. These scripts run in the webpage context as an extension to the DOM tree. Background pages, in contrast, are not loaded with each website; they react to browser events or carry out WebExtension API-based actions. Although content scripts have access to certain WebExtension API functions, their access is limited in scope. To leverage the full extent of the APIs, content scripts communicate with the background page via message passing.

While extensions can load static JavaScript as content scripts, they can also use a mix of host permissions and browser APIs to inject JavaScript into webpages programmatically. For example, an extension can request no websites under content scripts but then request \texttt{scripting} and host permissions on all websites to inject content scripts on websites dynamically. Furthermore, content scripts, without host permissions, must comply with website-defined cross-origin restrictions, unlike scripts injected via host permissions and API. This compliance limits their interaction with external entities, although they can still send and receive messages from the extension's background script.

\section{Security Landscape of Extensions}
\label{sec:security-analysis}

We conduct a comprehensive security analysis to understand potential design issues in the accessibility of input fields by extensions in Google Chrome, Mozilla Firefox, and Safari browsers. We focus on these browsers due to their widespread popularity and usage. We introduce a practical threat model and a systematic methodology for evaluating the potential risks posed by malicious extensions. We note here that while JavaScript running on the page can also access the HTML elements in a similar way, we restrict our analysis to extensions in this work as they operate within a controlled environment constrained by browsers' policies which allows us to identify and analyze potential security risks. 

\subsection{Extension Priviledges}
\label{subsec:extension-priviledges}
An extension can request different permissions to perform its functionalities. The permission model for browsers is largely static i.e. users agree to an extension's host and API permissions at install time, granting the extension a pre-defined set of capabilities. The user can also approve the websites where the extension can inject content scripts at install time. However, through extension settings, the browser provides a high-level control that allows users to prevent extensions from injecting scripts on certain websites.

\begin{figure}[t]
    \centering
    \includegraphics[width=0.85\columnwidth]{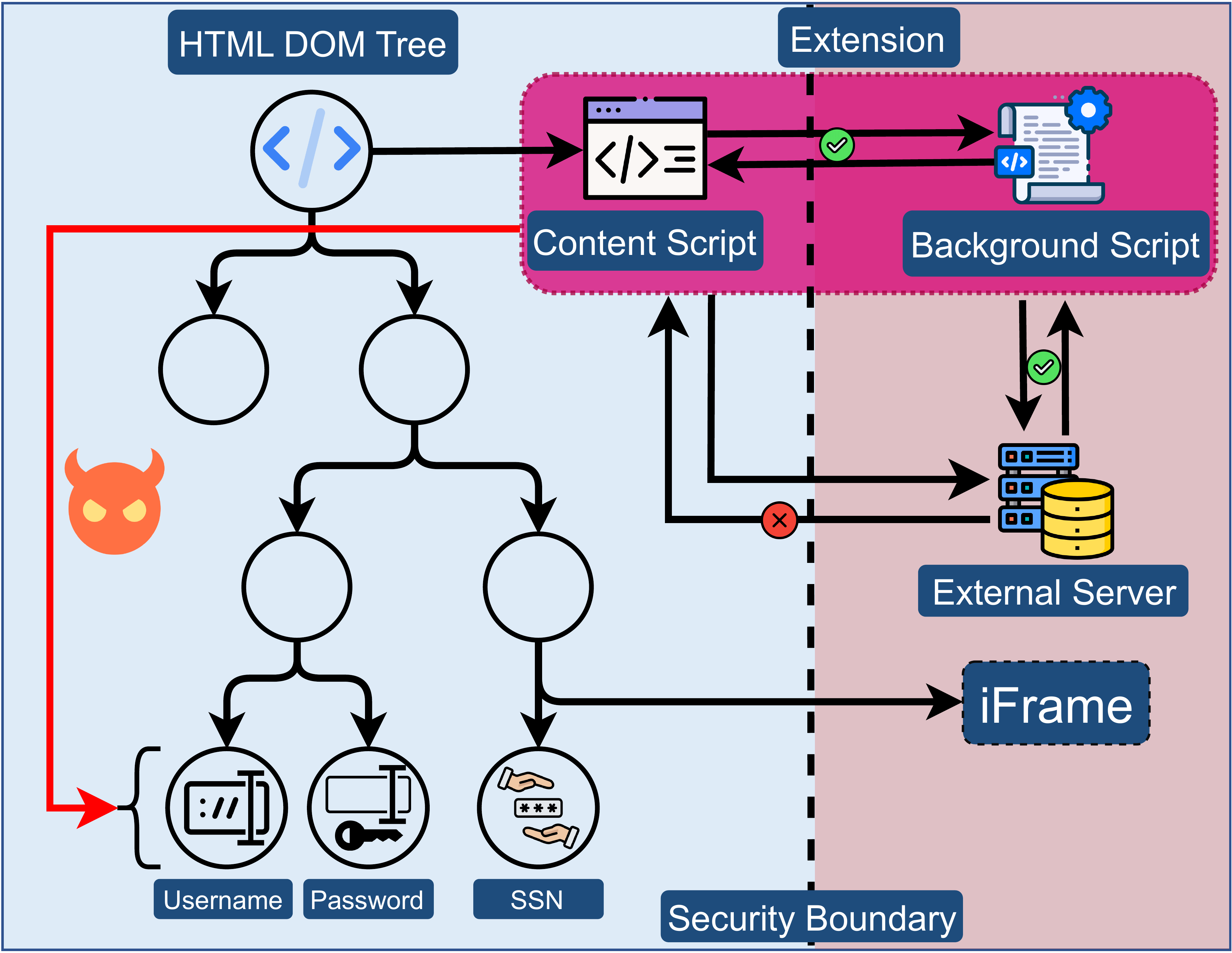}
    \caption{The figure illustrates the different contexts in which extensions and iframes exist with respect to a webpage. An extension's content script exists in the same context as the webpage, while an iframe is separated from the webpage by the website's Content Security Policy (CSP). Additionally, the content script can access any element of the DOM tree, including sensitive user data}
    \label{fig:attack_pipeline}
\end{figure}

The existing permissions framework across all browsers exhibits a coarse-grained approach, particularly with respect to access to web page content. The interaction of extensions with the HTML DOM tree is shown in \Cref{fig:attack_pipeline}. Once an extension is loaded on a webpage, it has unrestricted access to all elements on the page, including sensitive input fields. Such an extension, essentially a JavaScript program loaded into the DOM tree of the page, can access and potentially manipulate any data in the input fields on the page (\Cref{fig:attack_pipeline}). This coarse-grained control contrasts with the fine-grained access control for certain software and hardware resources, such as location information or file storage. 

One consequence of this coarse-grained model is the absence of a security boundary between the extension and the HTML elements (\Cref{fig:attack_pipeline}). This contrasts iframes governed by strict cross-origin policies that restrict access to the parent DOM tree and thus lie outside the security boundary. We examine this coarse-grained permission system in detail in \Cref{sec:attack-design} and show that it has security design issues that can violate security design principles and lead to unwanted access by extensions.

\noindent
\textbf{Isolating Extensions.}
Browsers follow a set of rules to isolate extensions to their own environment. Before December 2020, Manifest V2 (MV2) governed the extensions' interactions within the browser's boundaries. Previous research~\cite{} identified MV2's limitations and showed how extensions can bypass it, raising security issues. One significant MV2 security loophole was allowing \texttt{eval()} statements, enabling extensions to execute any external JavaScript without any checks. This lead to attacks such as iframe-based phishing and password stealing~\cite{4_perrotta_botnet_2018}.

\subsection{Existing Threats} 
\label{subsec:existing-threats}

Over the last decade, researchers have described various attacks employed by malicious extensions to extract sensitive data from input fields~\cite{20_Wang_firefox_2012,6_obimbo_analysis_2018,7_eriksson_hardening_2022,26_Varshney_detecting_2017}. These attacks can be broadly classified into static and dynamic code-based attacks. A summary of these works is shown in \Cref{l1}.

\noindent
\textbf{Static Code Attacks:} In these attacks, the malicious code is statically present in the extensions' source code. Prior works~\cite{20_Wang_firefox_2012,6_obimbo_analysis_2018,26_Varshney_detecting_2017} have proposed numerous attack vectors to exploit access of extension to the webpage and extract private information. For instance, Varshney et al.~\cite{26_Varshney_detecting_2017} have added an \texttt{eventListener} to log all the keystrokes. Similarly, Eriksson et al.~\cite{7_eriksson_hardening_2022} modify the password field to reveal the password values. We note that the underlying cause for these attacks is the lack of security boundary between the extension and the webpage, as discussed in ~\Cref{subsec:extension-priviledges}.

While these attacks are theoretically possible, they are also known to be impractical as they can be detected via static code analysis~\cite{23_Guha_verified_2011,26_Varshney_detecting_2017,11_zhao_automatic_2015,13_wang_combined_2018,12_dekoven_malicious_nodate}. This is evident because none of the existing works performing static code attacks have submitted the extensions to the web store. Therefore, these approaches do not test the viability of using extensions to compromise actual users that install extensions via web stores. 

\begin{figure}[H]
\begin{lstlisting}[style=htmlcssjs]
...
$(document).keypress(function(e) {
  ...
  keyHistory += String.fromCharCode(e.which);
  ...
});
...
\end{lstlisting}
\caption{Static Code Example}
\label{fig:static_code}
\end{figure}

\noindent
\textbf{Dynamic Code Attacks:} Dynamic code attacks involve retrieving the malicious code from an external server and then executing it into the target webpage~\cite{4_perrotta_botnet_2018, 20_Wang_firefox_2012, 6_obimbo_analysis_2018}. For example, Perrotta et al.~\cite{4_perrotta_botnet_2018} inject malicious remote code to perform a phishing attack. 
\begin{figure}[H]
\begin{lstlisting}[style=htmlcssjs]
...
fetch('evil_server_url')
  .then(response => response.text())
  .then(maliciousCode => {
    eval(maliciousCode); 
  });
...
\end{lstlisting}
\caption{Dynamic Code Example}
\label{fig:dynamic_code}
\end{figure}
\Cref{fig:dynamic_code} shows a skeleton code snippet where a malicious code hosted at a remote server is executed using the \texttt{eval()} function, which allows for the execution of JavaScript code represented as strings. These attacks can also include obfuscated code to prevent detection against dynamic analysis, such as remaining dormant and capturing data after a certain interval. 

Dynamic code attacks primarily exploit the remote code execution privileges provided to the extension. The remote code can not be vetted using code analysis methods, and thus are harder to detect as the injected code cannot be analyzed during the review process~\cite{}. To address this vulnerability, Chrome recently introduced new regulations that ban remote code execution. We discuss this in \Cref{subsec:sec-manifest-v3}.

\subsection{Security landscape after Manifest V3}
\label{subsec:sec-manifest-v3}
In December 2020, Google Chrome introduced Manifest V3 (MV3), bringing substantial changes in privacy, security, and performance. From a security standpoint, MV3 introduced \texttt{declarativeNetRequest} API for network request modifications and discontinued the \texttt{webRequest} API, disallowing extensions to modify network requests in real-time, closing a major loophole~\cite{google_mv3_overview}. MV3 also prohibited the execution of remotely hosted code and the use of eval statements. This vulnerability was exploited by attackers~\cite{4_perrotta_botnet_2018} to extract sensitive user data, as discussed in \Cref{subsec:existing-threats}.

Despite MV3's intended advancements in user privacy and security, content scripts' operations remain unchanged. This maintains the lack of security boundary between the extension and web page and  allows an extension to be loaded on the DOM tree and gain unrestricted access to the webpage, posing security risks for the users. We use this vulnerability to design our attack in \Cref{sec:attack-design}.

\noindent
\textbf{Adoption of MV3.}
Following Google's lead, Mozilla updated its browser ecosystem to MV3. By May 2023, Safari, Firefox, and all Chromium-based browsers, such as Microsoft Edge, had integrated support for MV3. However, only Chromium browsers have stopped accepting MV2 in their respective web stores, with Safari and Firefox continuing to allow MV2-based extensions.

\noindent
\textbf{Impact on Review Process.}
Before MV3, Chrome's web store review process involved using malicious extension-detecting systems like WebEval~\cite{Jagpal_2015}. WebEval utilized both static and dynamic analysis along with developer-centric heuristics (like email, age, etc.) to determine if an uploaded extension was malicious. However, as demonstrated by~\cite{4_perrotta_botnet_2018}, extensions can bypass this system, enabling the successful upload of a malicious extension to the web store. After MV3, Google prohibited all remote code execution and mandated that all code be included within extensions as this permits more reliable and efficient reviews of extensions submitted to the web store~\cite{google_mv3_overview}.

%


\subsection{Threat Model}
\label{subsec:threat-model}



In our threat model targeting browser extensions, we assume that the adversary adheres strictly to the guidelines provided in Chrome's MV3, and does not rely on side-channel attacks. Consequently, our assumption restricts the analysis to Chrome and Chromium-based browsers. This is a reasonable assumption because MV3 provides the latest framework, with other browsers adopting it as the standard framework (\Cref{subsec:sec-manifest-v3}). We also assume that the extension is uploaded to the webstore, and passes the Chrome review process. All changes made to the extension go through the review process of the webstore. For example, if an adversary acquires an existing extension and adds malicious code while updating the extension, the updates are passed through the review system. 
Finally, we assume that users directly download and install these extensions from the Chrome Web Store, as opposed to third-party installations. This ensures that the attack is practical, bypasses Google Chrome's review process, and the extension is available to users via the website, Once installed, users interact with these extensions in their regular browsing environment, unaware of the underlying threat.


\subsection{Security Analysis Methodology}
\label{subsec:security-analysis-methodology}

We conduct an experimental security analysis on browsers to examine how malicious extensions (as defined by our threat model) can 
potentially exploit the input fields. Specifically, we look for potential vulnerabilities in the latest browser extension regulation, MV3, and analyze if an extension with minimal permissions under MV3 can access sensitive input fields. Should such a possibility exist, we further explore if and how these extensions might access the DOM without the need for concealed code injections. With the attack, we discuss how it stems from the security design issues in the permission system, how it violates the security principles, and how it can lead to the exposure of sensitive information. We present our findings in \Cref{sec:attack-design}.

To evaluate the \textit{practicality} of extensions exploiting the vulnerabilities, we construct a proof-of-concept malicious extension to extract login credentials on websites. The extension requires enabling the content script to run on all web pages. Once installed, this extension accesses the input fields of login pages and extracts the username and password. We then submit the extension to Chrome Webstore where it bypasses the security checks and the review process. Finally, we run this extension on the top 10K domains from Tranco and extract login details. We discuss the measurement study in \Cref{subsec:website}. We also perform case studies to understand the implications of the vulnerability on input fields consisting of SSNs and Credit Card information. We discuss these results in \Cref{sec:case_studies}.

We measure \textit{prevalence} of malicious extensions in two ways. First, we analyze the extensions' potential ability to access sensitive input fields. We collect the requested permissions of the extensions and examine how many apps have the necessary permissions to extract sensitive information. Second, we perform static and dynamic analyses of extensions to identify if the extension is actually accessing the input fields. We present a detailed analysis of this measurement in \Cref{subsec:extensions-vulnerability}. It's important to note here, that this analysis does not aim to identify malicious extensions but to highlight that extensions currently have unrestricted access to sensitive fields, thereby potentially exposing sensitive data.

\begin{table*}[!ht]
    \centering
    \footnotesize
    \begin{tabularx}{\textwidth}{
            >{\centering\arraybackslash}m{1.2cm}
            >{\centering\arraybackslash}m{1.4cm}
            >{\centering\arraybackslash}m{3cm}
            >{\centering\arraybackslash}m{2.6cm}
            >{\centering\arraybackslash}m{2.9cm}
            >{\centering\arraybackslash}m{1.5cm}
            >{\centering\arraybackslash}m{1.2cm}}
    \toprule
        \textbf{Paper} & \textbf{Attack} & \textbf{Summary} & \textbf{Permissions} & \textbf{Content Script} & \textbf{Uploaded to WebStore} & \textbf{Type of Attack} \\ \midrule
         & Email Spamming & Snoop on users' legitimate email credentials to send them spam emails. & tabs, http://*/*, https://*/* & http://*/*, https://*/* & \faTimes & Static \\ \cmidrule{2-7}
        Liu et al.~\cite{1_Liu2012ChromeET} & DDoS Attack & Perform cross-site HTTPS request to launch DDoS attack on a victim webpage & tabs, http://*.yahoo.com/* & http://*.yahoo .com/* & \faTimes & Static \\ \cmidrule{2-7}
         & Password Sniffing & Steal user's email ID and password from login pages & tabs, https://online. citibank.com/* & https://online. citibank.com/* & \faTimes & Static \\ \midrule
        Bauer et al.~\cite{8_bauer_analyzing_2014} & iframe-based Password Stealing & Load a victim page in an iFrame and steal auto-filled credentials. & http://*/*, all\_frames, webRequestBlocking & http://*/*, https://*/* & \faTimes & Static \\ \midrule
        Perrotta et al.~\cite{4_perrotta_botnet_2018} & iframe-based Phishing & Inject malicious remote code into a website and perform a phishing attack. & N/A & http://*/*, https://*/* & \faCheck & Dynamic (uses eval) \\ \midrule
        ~ & Keylogger & Log all keystrokes and send them via SMS & http://*/*, https://*/* & http://*/*, https://*/* & \faTimes & Static \\ \cmidrule{2-7}
        \multirow{3}{*}{\parbox{1.1cm}{Varshney et al.~\cite{26_Varshney_detecting_2017}]}} & Credential Stealing & Automatically record autofilled user login info  & http://*/*, https://*/* & http://*/*, https://*/* & \faTimes & Static \\ \cmidrule{2-7}
         & Phishing & Stealthily redirect victim to a target website to perform phishing attack & tabs, http://*/*, https://*/* & Uses background scripts in conjunction with tabs API & \faTimes & Static \\ \cmidrule{2-7}
        ~ & Email Spying & Malicious Extension's content script reads the content of emails  & N/A & http://mail.google.com, https://mail.google.com & \faTimes & Static \\ \midrule
        Our Paper & Sensitive Data Sniffing & Read the value of a specific input field, as specified by a remote server. & None & <all\_urls> & \faCheck & Hybrid \\ \bottomrule
    \end{tabularx}
    \caption{Summarizing related works and their attacks against ours.}
    \label{l1}
\end{table*}

\section{Attack Design}
\label{sec:attack-design}

We analyze the interaction of browser extensions with sensitive input fields to identify potential vulnerabilities. We then design attacks to exploit the observed vulnerabilities and develop a proof-of-concept extension capable of extracting sensitive information within our threat model. We also upload our extension to the web store, bypassing the security measures in place during the review process. Finally, we analyze the root causes of attack success, mapping the vulnerabilities to violations of security principles. 

\subsection{Identifying Vulnerabilities}
\label{subsec:basic-principles}
%

\begin{figure}[t]
    \centering    
    \begin{subfigure}[b]{0.48\columnwidth}
        \centering
        \includegraphics[width=\textwidth]{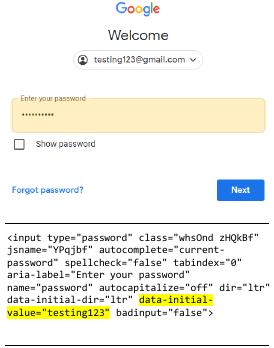}
        \caption{\textit{Type-A} vulnerability on google.com}
        \label{fig:google_typ_1}
    \end{subfigure}
    \hfill
    \begin{subfigure}[b]{0.48\columnwidth}
        \centering
        \includegraphics[width=\textwidth]{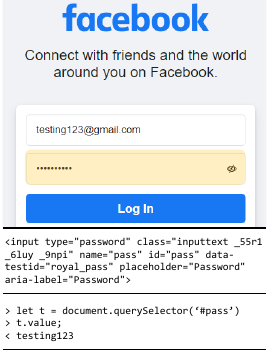}
        \caption{\textit{Type-B} vulnerability on facebook.com}
        \label{fig:facebook_typ_2}
    \end{subfigure}
    \caption{Different types of vulnerabilities found in the wild. (a) The vulnerability allows a malicious extension attached to the DOM tree to extract login credentials. (b) The password is visible in the \texttt{outerHTML} of the element and can be extracted directly from the source code.}
    \label{fig:vulnerabilities}
\end{figure}

As discussed in \Cref{sec:background}, when an extension is loaded onto a website, it is integrated into the DOM tree, obtaining unrestricted access to all DOM elements via the DOM APIs. This exposes a critical security issue -- the lack of a security boundary between the extension and the rest of the DOM tree, as shown in \Cref{fig:attack_pipeline}. Unlike iframes, which have isolated DOM trees, extensions face no restrictions once permitted to operate on a page, resulting in a broad attack surface.
From this design, we discover two vulnerabilities, labeled as \textit{Type-A} and \textit{Type-B}. Illustrations of websites suffering from these vulnerabilities can be found in \Cref{fig:vulnerabilities}.

In \textit{Type-A} vulnerability, the sensitive values are visible in plain text in the source code of the webpage. For example, in the case of password fields, the password values are present in the \texttt{outerHTML} of the element, as shown in \Cref{fig:google_typ_1}. This is particularly concerning as any entity with access to the source code, including the extensions, can extract the password values. This security lapse also has implications on possible defenses that we discuss in \Cref{sec:remedies}. We also note that, to the best of our knowledge, we are the first ones to highlight the \textit{Type-A} vulnerability.

In \textit{Type-B} vulnerability, the extension can gain access to the input elements via the DOM API, and extract the input element's value using the \texttt{.value} function. The values of sensitive input fields such as password fields are often masked to prevent shoulder surfing~\cite{lashkari2009shoulder} or screenshot attacks~\cite{7_eriksson_hardening_2022}. This masking is done either via HTML input types (e.g. \texttt{type=password}) or through JavaScript-based obfuscation methods (e.g. private variables using WeakMaps\footnote{\url{https://modernweb.com/private-variables-in-javascript-with-es6-weakmaps/}}). Notably, accessing the value using \texttt{.value} function bypasses the HTML-based masking, as shown in ~\Cref{fig:facebook_typ_2}.

\subsection{Proof-Of-Concept Extension}
\label{subsec:proof-of-concept-extension}
Prior work has exploited the lack of security boundary between the extension and the rest of the DOM tree~\cite{20_Wang_firefox_2012,6_obimbo_analysis_2018,7_eriksson_hardening_2022,9_carlini_evaluation_nodate,26_Varshney_detecting_2017}. They either used static or dynamic code injection to extract sensitive data. Extensions with static code are impractical as the malicious code can be detected via code analysis~\cite{23_Guha_verified_2011,26_Varshney_detecting_2017,11_zhao_automatic_2015,13_wang_combined_2018,12_dekoven_malicious_nodate}. On the other hand, dynamic code injection attacks are not feasible after the introduction of MV3. Thus, to build a practical extension and exploit the observed vulnerabilities, we need - a)  to access the input elements without using dynamic code injection; b) to overcome any obfuscation on the values of input fields; and c) to submit the extension to chrome web store and clear the review process.

We note here that a malicious extension can also manipulate the elements and modify the content to perform other attacks such as screenshot attack~\cite{7_eriksson_hardening_2022} and phishing attacks~\cite{8_bauer_analyzing_2014,4_perrotta_botnet_2018}. However, in this work, we focus on the security of text input fields and the sensitive information that can be extracted from them. We devise three attacks based on the observed vulnerabilities: \textit{source extraction attack}, where we copy the sensitive values of input fields from the element's \texttt{outerHTML}; \textit{value extraction attack}, where we select the target input field and read the sensitive values; and \textit{element substitution attack} where we bypass JavaScript based obfuscation to extract sensitive values.

Our primary objective is to build a practical extension to pass the webstore review process and extract sensitive information. To achieve this, we propose a hybrid attack that leverages techniques from static and dynamic code injections. Specifically, we design our extension to include a benign code template that identifies an element with a given \textit{CSS selector}. We dynamically retrieve the \textit{CSS selector} string from a server which allows us to control the input fields at runtime. This technique is similar to that used by Khandelwal et al.~\cite{cookieenforcer}. We do not require additional permission to communicate with the server and retrieve the \textit{CSS selector}. We instead use the background page to fetch the string and pass it through messages to the content script, as shown in \Cref{fig:attack_pipeline}.
Finally, we identify the type of attack to be used by analyzing the element's properties. A skeleton code of the extension is shown in \Cref{fig:skeleton_code}. Note that this is different than the code injection attacks as they obtain the code and execute them using \texttt{eval} statements. In our case, we only get the CSS selector string from the server, and we avoid using any eval statement. Additionally, our extension only requires minimal permission to run on the given page. Next, we describe the different attacks in detail.

\begin{figure}[h]
\begin{lstlisting}[style=htmlcssjs,label={example_code}]
...
fetch('server_url') // Retrieve CSS selector
  .then(response => response.text())
  .then(data => {
    var els = document.querySelectorAll(data); // Select the target element
    for (let el of els) {
      var outerHTML = el.outerHTML
      var typeA = checkForTypeA(outerHTML);  // Determine if Type-A 
      if (typeA){
          el.addEventListener(text, sourceExtractionScript)
      }
      else{
          el.addEventListener(text, valueExtractionScript)
  }}});
...
\end{lstlisting}
\caption{Skeleton code showing how the extension extracts the content of sensitive fields by determining the type of vulnerability that can be exploited.}
\label{fig:skeleton_code}
\end{figure}

\noindent
\textbf{Source Extraction Attack:} In this attack, we exploit the vulnerability where the sensitive values are present in the source code of the HTML. In this attack, we wait for the user to enter their credentials and capture the webpage's HTML when the user clicks the Login/Sign-in button, and the page is redirected. At this instant, we copy the page's HTML source code, using the \texttt{outerHTML} property of the \texttt{window.document.body} object. Once captured, we can use regex to capture only the \texttt{password} type input field and send it back to the server. This is a novel attack leveraging the security lapse found in many websites, including \url{google.com}, and can potentially impact billions of users, as discussed in \Cref{subsec:website}.

\noindent
 \textbf{Value Extraction Attack:} This attack involves identifying and selecting the target input field. Using JavaScript, this can be done in several ways, e.g. xpaths, CSS-selectors, etc. For our attack, we use the CSS selectors to identify the elements. Specifically, we use \texttt{document.querySelectorAll} method takes a CSS selector string and returns all the elements that match the selector. For example, for password fields, we use the selector \texttt{input[`type=password']} and gain access to all the input elements with the password tag. Note that such selectors can be generated for other sensitive fields too. As discussed in \Cref{subsec:basic-principles}, websites can use HTML-based or JavaScript-based obfuscation. With these obfuscation methods, the text entered is not visible on the screen. However, for HTML-based obfuscation, we can access the values using \texttt{.value} once we can access the elements.

 \noindent
 \textbf{Element Substitution Attack:} JavaScript-based obfuscation prevents access to the information using the \texttt{.value} function. For example, for SSN fields, \url{chase.com} uses JavaScript-based obfuscation, which prevents access to the values. To extract values from these elements, we propose \textit{Element Substitution} attack where the extension can leverage the lack of security boundary between the extension and the DOM tree and replace the protected input element with a simple password field. This new input field still appears secure, however, the values can be extracted using the \texttt{.value} function. We show a skeleton code corresponding to the element substitution attack in \Cref{fig:substitution}. 

\begin{figure}[h]
\begin{lstlisting}[style=htmlcssjs,label={example_code2}]
...
fetch('server_url')
  .then(response => response.json())
  .then(data => {
    var old_element = document.querySelector(data.selector);
    var new_element = document.createElement(data.tag);
    new_element.setAttribute('type', data.type);
    new_element.name = old_element.name;
    ...  // Add other attributes
    old_element.parentNode.replaceChild(new_element, old_element);
  });
...
\end{lstlisting}
\caption{Skeleton code showing the Element Substitution Attack, where an obfuscated input element is swapped for a basic text-input field.}
\label{fig:substitution}
\end{figure}



\subsection{Uploading to Web Store}
\label{subsec:uploading-to-web-store}
Finally, we submit the extension to the web store to evaluate the web store's review process. The extension passed the review process on Google Chrome web store. 
To hide the extension's malicious aspects, we disguised it as a GPT-based assistant offering ChatGPT-like functions on websites. The extension asked for permission to run on all websites, which is reasonable as most extensions that offer assisting features ask for this permission.

Webstores' failure to identify the malicious extension highlights the need for more robust verification systems for browser extensions. The existing security checks may not be sufficiently comprehensive or effective in identifying potential threats. This is particularly concerning given the potential for extensions to access sensitive user data, including passwords and other input field data, as shown in this work. 

\noindent
\textbf{Ethical Considerations.} We maintained ethical integrity throughout the process by adhering to the established guidelines from prior works~\cite{4_perrotta_botnet_2018}. Specifically, we ensure we do not collect sensitive information from manual testers during the review process. Our extension was engineered to interact with our servers, identify the type of HTMLElement we were targeting (in this case, input elements), monitor the values on those elements, and ultimately transmit the recorded values back to our server. To protect the privacy of the manual tester while not revealing the extension's malicious nature, we deactivated our data-receiving server, retaining only our element-targeting server online. Consequently, our extension would request the target element, acquire the CSS selector, and then attempt to send the recorded data to a non-existent server. This procedure ensured that the primary operation of the extension remained consistent with our original design. We uploaded the extension to the web store once, ensuring we did not waste testers' time during the manual review process. Additionally, once approved, we immediately removed the extension from the web store. We always kept extension in ``unpublished'' mode so the users could not find and install the extension.



\subsection{Root causes for attack success}
\label{subsec:root-cause-for-attack-success}


Next, we analyze the root causes for the success of our attacks. The success of our attack stems from a combination of factors, including the improper application of security principles, a trade-off between usability and security, bad practices by websites, and flaws in the online review process of extensions. In this section, we discuss these root causes and their implications for the security of sensitive input fields.
\smallskip

\noindent
\textbf{Improper Application of Security Principles.} The success of our attack can be largely attributed to the improper application of fundamental security principles within the current design of web browsers and extensions. One of the key issues lies in the coarse-grained permission model at the HTML level. Once an extension is allowed to run on a page, it has unrestricted access to all elements. This unrestricted access is in direct violation of the Principle of Least Privilege, a fundamental security principle that advocates for limiting the permissions granted to a process to only those that are necessary for its function.

This unrestricted access also undermines the principle of Complete Mediation, which requires that every access to a resource be checked for appropriate permissions. In the current model, once an extension has been granted access to a page, subsequent accesses to elements on the page are not checked, allowing the extension to interact with all elements on the page freely.


These violations of fundamental security principles create an environment where sensitive data is vulnerable to unauthorized access and manipulation, highlighting the need for a more secure design that adheres to these principles.

\smallskip

\noindent
\textbf{Trade-off Between Usability and Security.}
In security systems, there is often a trade-off between usability and security. This trade-off is clearly evident in the current security landscape of web browsers and extensions. Websites often rely on browsers to provide the necessary security protections, placing trust in the browser's ability to safeguard sensitive data. However, this trust can lead to vulnerabilities if the browser's protections are insufficient or can be circumvented by malicious extensions.

An example of this trade-off can be seen in password managers. While they aim to improve convenience by storing and automatically filling passwords, they require access to password fields, which can compromise security measures. This creates a challenge in balancing between implementing strict security measures and ensuring the smooth operation of password managers.

Interestingly, we observed that many websites attempt to obfuscate Social Security Numbers (SSNs) but not passwords. This suggests a recognition of the need for protection of sensitive data, but an inconsistent application of it. The decision to obfuscate SSNs but not passwords may be driven by a desire to balance usability and security, but it also highlights the complexities and potential pitfalls of this trade-off.

\smallskip

\noindent
\textbf{Websites' Bad Practices.} Our case studies revealed a range of security practices across different websites, with some leaving sensitive input fields unprotected or implementing only minimal protections. The reasons for these practices are not always clear, but they contribute to the overall vulnerability of these input fields. Even obfuscation, while better than leaving sensitive data in plain sight, is insufficient to fully secure these fields.

\smallskip

\noindent
\textbf{Flaws in Online Review Process of Extensions.} Lastly, the online review process for extensions, particularly those with dynamically loaded selectors, has significant flaws. These flaws can allow malicious extensions to pass through the review process undetected, providing them with a platform to launch attacks. It's important to note that this is different from code injection, as the malicious code is part of the extension itself, not injected into the webpage.

These factors combine to create a landscape where sensitive input fields are vulnerable to attack. In \Cref{sec:measurements} and \Cref{sec:case_studies}, we measure the \textit{practicality} and \textit{prevalence} of this vulnerability on real websites by conducting large-scale measurements (\Cref{sec:measurements}) and case studies (\Cref{sec:case_studies}).
\section{Measurements}
\label{sec:measurements}
To evaluate the \textit{practicality} of the vulnerability of sensitive text input fields, we analyze the login pages for vulnerabilities from the top 10K websites from Tranco\footnote{List available here: \url{https://tranco-list.eu/list/4K4GX/1000000}}. We take password fields as a representative for sensitive input fields as we can automatically detect password fields, allowing us to scale the analysis. We analyze other sensitive input fields through case studies in \Cref{sec:case_studies}. To measure the prevalence of potentially malicious extensions, we first examine the extensions' permission to understand their \textit{potential ability} to access password fields. We then perform static and dynamic analysis to identify extensions accessing password fields using the vulnerabilities discussed in \Cref{sec:security-analysis}.

\begin{figure}[ht]
    \centering
    \includegraphics[width=\columnwidth]{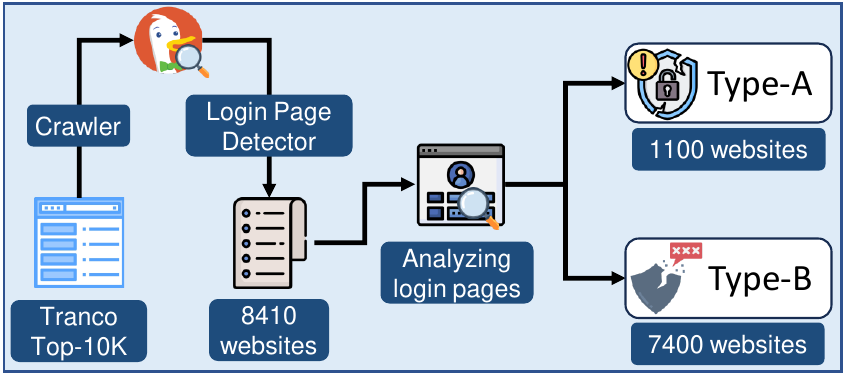}
    \caption{Our website vulnerability measurement pipeline uses a custom crawler to identify login pages of websites and detect the type of vulnerabilities present.}
    \label{fig:website_vuln_methodology}
\end{figure}

\subsection{Websites' Vulnerabilities}
\label{subsec:website}

We conduct a comprehensive measurement of the security vulnerabilities associated with password fields. Our infrastructure consists of a custom-built web crawler to navigate popular websites' login pages and inspect the HTML and JavaScript (JS) elements associated with password fields. The crawler is equipped with capabilities to handle different types of login forms, including both static and dynamic forms. It is also designed to detect and categorize the two types of vulnerabilities: Type-A, where password values are visible in plain text in the HTML source code, and Type-B, where password values are obscured but can be accessed via JavaScript. We ran the crawler from a controlled environment to ensure consistency in the measurements. 

\noindent
\textbf{\textit{Methodology:}}
 We perform the measurement using a Chromium browser controlled via the Selenium library in Python. We also install the extension that performs the attack (\Cref{sec:attack-design}) to extract the passwords. The overview of the measurement pipeline is shown in \Cref{fig:website_vuln_methodology}.

We use the top-10K domains from the Tranco list generated on Feb. 2nd, 2023. We employ a two-tiered approach to identify and analyze the login pages of these domains. First, we attempt to locate the login button on the homepage of each domain by analyzing the text of all clickable elements on the page and searching for keywords associated with the login function. In case of failure, we perform a search on DuckDuckGo using the query \texttt{<domain name> log[-?]in}. We then select the top five pages from the search results as potential login page candidates and analyze each candidate page to determine if it is a login page. In particular, we treat a page as a login page if there is a \textit{username/email} field or a \textit{password} field. 

After finding the login page, we automatically enter a unique username and password and attempt to extract them using the extension. We note that a login page can exist without password fields (e.g., \url{linkedin.com}). Specifically, there can be login pages where the password fields appear only after the email/username is entered. To capture the password field in such cases, we press ENTER after inserting the username and check if the password field is present. This allows us to capture login pages where the password fields are initially hidden.

\noindent
\textbf{\textit{Results:}} In our study, we identified login pages for 8,410 websites out of the top 10,000 domains. Among these, we found password fields present on 7,140 websites. The remaining 1,270 pages contained username or email fields but no password fields. We manually inspected a subset of these web pages to investigate the causes. We discovered that certain websites do not display password fields unless the provided email address corresponds to a registered account. For instance, on \url{quip.com}, the page redirects to third-party sign-in portals. Additionally, we came across cases where non-login web pages contained an email field, leading to false positives in our login page detection. It's important to clarify that these false positives do not affect the results presented in this work, as these pages lacked the password fields.

Notably, we could extract password data from all the websites that presented the password fields. Further analysis revealed that 1,100 websites exhibited \textit{Type-A} vulnerability; the password values were displayed in plain text within the HTML DOM. \Cref{fig:vulnerabilities} shows snapshots of these vulnerabilities, depicting password values in plain text in the HTML. The underlying issue is that the \textit{value} attribute of the \textit{input} element is set to update at each keystroke. In most implementations of password fields, this value attribute is omitted or kept empty. 

Alarmingly, we find that the \textit{Type-A} vulnerability was present on several popular websites, including but not limited to \url{gmail.com} and \url{cloudflare.com}. Gmail has over 1.8B monthly active users, whereas Cloudflare, a cybersecurity company, has over 4M active users. The results indicate that this security vulnerability can potentially impact billions of users. It is noteworthy that cloudflare.com is one of the popular cybersecurity firms used by services like OpenAI, Google, etc., for cybersecurity needs. The existence of such a basic security oversight exists on popular websites is concerning, as even websites with substantial resources are not immune to security lapses. This puts the user at significant risk as any JS code can access the HTML, send it to a private server, and extract passwords without running a JS code. We further discuss the implication of \textit{Type-A} vulnerability on possible defenses in \Cref{sec:remedies}.


\begin{figure}[ht]
    \centering
    \includegraphics[width=\columnwidth]{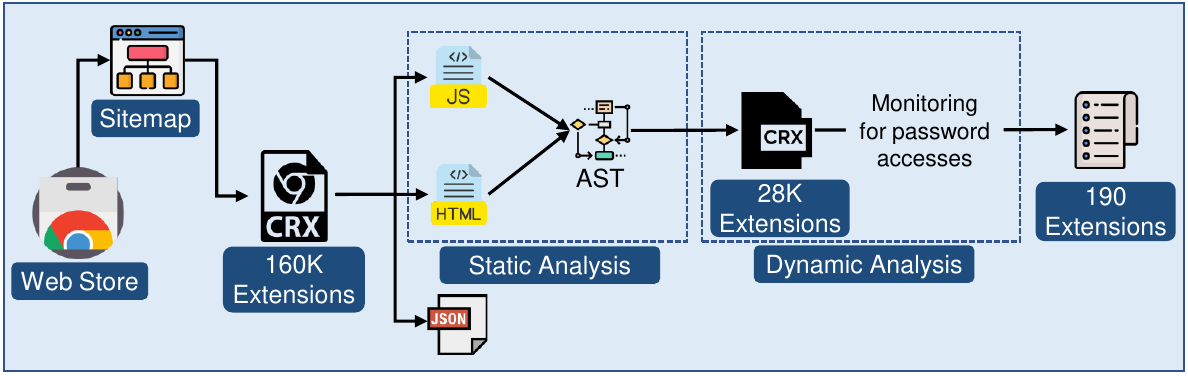}
    \caption{Our extension analysis pipeline uses a mix of static analysis, filtering out extensions that select input fields, and dynamic analysis to check if the password field's content is stored.}
    \label{fig:extension_lint_flow}
\end{figure}
\subsection{Extensions Vulnerability}
\label{subsec:extensions-vulnerability}

\subsubsection{Potential Ability To Exploit Vulnerability}
 We analyze the extensions on the Chrome store to identify how many extensions can potentially exploit the vulnerabilities discussed in \Cref{sec:attack-design}. We first download the available extensions by scraping the Chrome web store. The downloaded extensions include the manifest.json file, HTML files, JavaScript files, and image files associated with the extension. We then analyze the manifest files and look for extensions that request the \texttt{scripting} permission, or that request the content scripts to be run on \texttt{all\_urls}. Depending on the use case, an extension can request content scripts that have access to the HTML DOM tree to be run on specific web pages or all web pages. \texttt{Scripting} permission allows the extension to inject content script.
 
 Analyzing the manifest files, we find that  12.5\% (17.3K) extensions have the necessary permissions to extract sensitive information on all web pages. This includes popular extensions such as \textit{AdBlockPlus} and \textit{Honey} with more than 10M users. We also find that 33.6\% (46.4K) extension request content scripts to be run on at least one website, whereas 55.2\% (76.3K) extensions do not request \texttt{scripting} or content script permissions. 


\subsubsection{Potential Prevalence}
Prior research has demonstrated the existence of malicious extensions in the webstore~\cite{1_Liu2012ChromeET,13_wang_combined_2018,7_eriksson_hardening_2022}. 
In this study, we focus on the potential for extensions to select and store password fields in a variable and aim to measure how many extensions access the password fields.

\noindent
\textbf{\textit{Methodology:}}
\Cref{fig:extension_lint_flow} shows the extension analysis pipeline. Our objective is to identify extensions that select any password fields. Identifying access to input fields is a challenging problem as JavaScript provides numerous methods to select a \texttt{HTMLInputElement}. Thus, filtering extensions using all possible selection methods is infeasible~\cite{24_louw_enhancing_2008}. Therefore, we perform static analysis and create custom ESLint rules to filter extensions that include a function containing the \texttt{querySelector} or \texttt{getElement} keywords and include \texttt{input} as its function parameter, as shown in \Cref{lst:eslint}. This selects extensions that are selecting input fields. This filtered list contains some extensions that do not perform any input field selection, but their function call matches our filtering criteria. Conversely, our filters may fail to capture extensions that use alternative forms of element selection.


\begin{figure}[t]
\begin{lstlisting}[style=htmlcssjs]
"CallExpression[callee.type='MemberExpression'][callee.property.name=/querySelector/] > Literal": function(node){
  if (node.value.toString().toLowerCase().includes("input")) {
    context.report({
      node: node,
      message: "Found input string"
    });
  }
}
\end{lstlisting}
\caption{ESLint code snippet to detect the use of \texttt{querySelector} to select elements based on CSS selectors having `input' as a substring.}
\label{lst:eslint}
\end{figure}

Next, we perform dynamic analysis to identify extensions that select and store password-type input fields. Following prior works~\cite{7_eriksson_hardening_2022, 11_zhao_automatic_2015}, we instrument the extension to determine whether the passwords are stored in a variable within the extension code. Specifically, we insert a \texttt{console.log} below the variable declaration to print its value.


% ...


% ...

Upon instrumenting the filtered set of extensions, we recompress them into CRX files and then use Selenium to load them automatically into a Google Chrome instance. We then visit the login pages of Facebook and Citi Bank, input a unique string in the username and password field, and verify whether these strings appear in the console window. If they do, we flag the extension as selecting and storing password-type input fields in variables.

\noindent
\textbf{\textit{Results:}}
Our scraping of the web store resulted in 160K extensions. After applying our static analysis filters, we retained 28K extensions. Dynamic analysis of these 28K extensions flagged 190 extensions storing password values in a variable. Of these 190 extensions, 12 had more than 10K downloads, and three had more than 100K downloads. While some flagged extensions functioned as password managers, many were random extensions that selected and stored password fields. For example, Remote Torrent Adder's extension, with over 40K downloads, accesses input fields and stores them in a variable. 

\subsection{Takeaways}
\label{subsec:takeaways}

\noindent
\textbf{{Systemic Issue.}}
Our measurement studies on the top 10K websites show that we could extract passwords from all the login pages with passwords. The widespread presence of these vulnerabilities indicates a systemic issue in the design and implementation of password fields. 

\noindent
\textbf{{Need for Stringent Security Measures.}}
The presence of \textit{Type-A} vulnerabilities, where passwords are visible in plain sight in the HTML source code, is concerning. This severe vulnerability bypasses any browser protections, even the ones presented in this paper (\Cref{sec:remedies}), leaving sensitive data exposed and easily accessible to anyone viewing the source code. This highlights the need for more awareness of security measures.

\noindent
\textbf{{Role of Password Managers.}}
The widespread use of password managers may partially explain the prevalence of \textit{Type-B} vulnerabilities, where password values are obscured but can be accessed via JavaScript. These tools enhance the user experience by automating the process of entering passwords, storing the encrypted passwords, and later auto-filling these fields when required~\cite{Zhao2013VulnerabilityAR_Passwords_1}. This functionality reduces the cognitive load on users and encourages the use of complex, unique passwords for each site, thereby enhancing overall security~\cite{Herley2012ARA_passwords_2}.

However, for password managers to function effectively, they require access to password fields via JavaScript. This necessity creates an inherent security vulnerability. While the password fields may appear obscured to users, any JavaScript code running on the page, including potentially malicious scripts, can access these fields and read their values. This interaction between password managers and \textit{Type-B} vulnerability presents a trade-off between usability and security. While password managers improve usability and promote better password practices, their operation necessitates JavaScript access to password fields that inherently creates a security risk. We discuss potential solutions to maintain usability while providing security in \Cref{sec:remedies}.

\subsection{Limitations}
\label{subsec:limitations}
\smallskip

\noindent
\textbf{Website Measurements.} We note two main limitations associated with our methodology. First, we may have missed dynamically loaded pages that rely on user interaction to reveal login forms. Second, our method for identifying login pages relied on the presence of certain HTML input fields (such as email and password fields). However, some websites may employ unconventional methods or unique identifiers for their login procedures, making it difficult to identify all login pages correctly.

\smallskip

\noindent
\textbf{Extension Analysis:} In our extension analysis, we use a combination of static and dynamic components to identify problematic extensions. During the static analysis phase, we only include extensions that select input fields with methods like \texttt{querySelector}, \texttt{querySelectorAll}, \texttt{getElementBy}, and \texttt{getElementsBy}. However, our static analysis can't include every extension that selects input fields due to the numerous ways to select elements.

In dynamic analysis, we modify the extensions to automatically insert a log statement into the variable holding a selected element. This lets us track extensions that store input data in a variable but misses extensions that process the input data directly without storage. Some malicious extensions activate after a time delay, which our method also misses. Finally, our dynamic analysis does not detect extensions that add an event listener to input fields instead of simply reading the values.

\section{Case Studies}
\label{sec:case_studies}
Text input fields like Social Security Numbers (SSNs) and Credit Card details may expose private data. Misuse of this data could lead to identity theft or financial fraud. We study the possible leakage of such data via input fields. We note that carrying out large-scale investigations to find such leaks is difficult, as this data is typically input after the user logs in.
Therefore,
we opt for a case study approach. In particular, we focus on popular websites where such information is commonly entered, aiming to understand how websites protect input fields. In this section, we present our findings from these case studies, providing insights into the current state of security for SSN and credit card input fields on the web.

\subsection{SSN Leakage}
\label{subsec:ssn_leakage}
Social Security Numbers (SSNs) are unique identifiers assigned to individuals in the United States for identity verification, tracking earnings, and credit checks. Unauthorized access or exposure of the SSNs of an individual can result in fraudulent loans or criminal charges being falsely attributed to the victim. According to the U.S. Department of Justice, millions of Americans fall victim to identity theft yearly, resulting in significant financial loss~\cite{harrell_victims_2018}. As more financial institutions require users to enter their SSNs for various transactions (e.g., tax returns or credit checks), the security of these input fields becomes increasingly critical. We examine websites that implement protections for SSNs and those that do not.

\noindent
\textbf{Websites Protecting SSNs.} Several financial institutions, including Chase, Wells Fargo Bank, and Bank of America, have implemented measures to protect SSN input fields. These measures involve using JavaScript-based obfuscation to protect the values entered into these fields. When a user enters their SSN on these websites, the input field displays obfuscated characters (such as asterisks or dots) instead of the actual SSN. This means that the extension can only see the obfuscated characters, providing a layer of protection against malicious extensions.

However, our research indicates that even these protections are not foolproof. Our element substitution attack, discussed in \Cref{sec:attack-design}, can bypass the security measures and access the information entered by substituting the HTML element containing the obfuscated SSN. This demonstrates that while obfuscation can provide a level of protection, it is not sufficient to secure sensitive input fields fully.

\smallskip

\noindent
\textbf{Websites Not Protecting SSNs.} On the other hand, websites such as IRS.gov, Capital One, and USENIX security do not implement any protections. In these cases, SSNs are visible in plain text. This presents a significant security risk, as any malicious extension could potentially access and steal this sensitive information. Furthermore, the lack of protection on these websites is particularly concerning, given their nature. For instance, IRS.gov is the official website of the U.S. Internal Revenue Service, and Capital One is a major financial institution.

These case studies highlight the variability in security measures implemented to protect SSN information across different websites. While some websites have protections in place, others exhibit significant vulnerabilities that could be exploited to gain unauthorized access to sensitive data.




\subsection{Credit Card Information Leakage}
\label{sec:cc_leakage}


Credit card information is another type of sensitive data frequently entered into website input fields. Unauthorized access to this information can lead to financial fraud, such as unauthorized transactions. Furthermore, credit card information is often used without additional authorization (such as two-factor authentication), making it a particularly attractive target for attackers. Here, we conduct case studies on a selection of websites to understand how they handle the security of credit card information in input fields.

\smallskip

\noindent
\textbf{Websites Protecting Credit Card Information.} One example of a website that implements measures to protect credit card input fields is \url{silvercar.com}. They add the attribute \texttt{type=password} to their credit card input fields. This obfuscates the information entered, similar to password fields, providing a layer of protection against unauthorized access. However, this protection is relatively weak, and our attack (\Cref{sec:attack-design}) can bypass this and access the information entered by calling the \texttt{.value} method on the input element. 


\noindent
\textbf{Websites Not Protecting Credit Card Information.} On the other hand, major online marketplaces such as Google and Amazon do not implement any protections for credit card input fields. In these cases, credit card details, including the Security Code and zip code, are visible in plain text on the webpage. This presents a significant security risk, as any malicious extension could potentially access and steal this sensitive information. The lack of protection on these websites is particularly concerning, given their scale and the volume of transactions they handle daily. 


\subsection{Takeaways}



Our case studies on the handling of sensitive information in input fields across various websites have yielded several key insights. We observed different security practices, from unprotected input fields to fields with minimal protections or JavaScript-based obfuscation. 

\smallskip

\noindent
\textbf{Unprotected Input Fields.} In the absence of any protective measures, as seen on websites like IRS.gov, Capital One, USENIX, Google, and Amazon, sensitive data such as SSNs and credit card information are immediately accessible to all extensions running on the page. This presents a significant security risk, as private data is left vulnerable.

\smallskip

\noindent
\textbf{Minimal Protection.} Some websites, like \url{silvercar.com}, attempt to protect sensitive input fields by using the \texttt{type=password} attribute. While this visually obfuscates the entered data, it only offers minimal protection. Value extraction attack described in \Cref{subsec:proof-of-concept-extension} can bypass this protection.

\smallskip

\noindent
\textbf{JavaScript Based Obfuscation.} A more sophisticated approach is seen on websites like Chase, Wells Fargo Bank, and Bank of America, where JavaScript-based obfuscation is used to protect the input elements. However, even this measure is not foolproof. Our \textit{element substitution} attack, which involves substituting the protected HTML element containing the sensitive information, can effectively circumvent this protection.

\smallskip

The success of our attack across all these scenarios highlights the fundamental issue: the extension's unrestricted access to all HTML elements on a page and the ability for them to manipulate these elements. This unrestricted access allows a malicious extension to bypass even the most sophisticated protections currently in place, highlighting the urgent need for more robust and comprehensive security measures. Notably, third-party JavaScripts loaded on the webpage are also attached to the DOM tree and have the same privileges as a browser extension. Therefore, the findings discussed here also apply to them, further exacerbating the security vulnerabilities of sensitive input fields. 
\section{Remedies}
\label{sec:remedies}
As we have shown in this work, the lack of security boundary between the extension and the webpage can allow a malicious extension to extract sensitive user information entered in input fields. In this section, we propose a two-fold approach to address these vulnerabilities. The \textit{Bolt-on} solution serves as an add-on package that web developers can use to secure sensitive input fields. In contrast, the \textit{Built-in} solution proposes a more fundamental browser-level solution that involves instrumenting the browser to alert users when sensitive fields are accessed.



\subsection{Bolt On}
\label{subsec:bolt-on}
In the \textit{Bolt-on} solution, we provide a JavaScript package that the developers can use to protect sensitive input fields. Specifically, we introduce a new \texttt{HTMLInputElement} type, \texttt{SecureInput}\footnote{\url{https://osf.io/nbdfj/?view_only=c496010851314a3299c9e816804aac52}} that leverages WeakMaps to store the sensitive information as private data.
Unlike previous solutions~\cite{1_Liu2012ChromeET,7_eriksson_hardening_2022}, our solution is ready to use and does not necessitate a major revamp of the current browser extension architecture.
Developers can simply import the secure-input library and designate any input they wish to secure as follows:
\begin{lstlisting}[style=htmlcssjs,label={lst:secure_input_example}]
<input is="secure-input" type="password">
\end{lstlisting}

The \texttt{SecureInput} class inherits all the properties associated with the base HTMLInputElement or the input tag.
We store the real value of the input field in the WeakMap while presenting a masked value to the value attribute of the \texttt{HTMLInputElement}. This approach effectively prevents \textit{Type-B} leaks (\Cref{subsec:basic-principles}). We note that the website retains full access to the input field and its methods as the \texttt{SecureInput} class is employed by the website. 

Furthermore, using WeakMaps ensures that most known attacks fail to access sensitive information. For example, Eriksson et al.'s~\cite{7_eriksson_hardening_2022} methodology of changing the input type to text would only display the masked value. Additionally, the \texttt{secure-input} library employs a mutation observer that alerts the user in case the secure-input field is switched with any other input field, by an attacker or otherwise, protecting against element substitution attacks (\Cref{sec:attack-design}).

We provide out-of-the-box support for authentication using \texttt{SecureInput} field. Specifically, we add a new \texttt{submit()} function that locates the username and communicates with the developer-specified server, returning the authentication status. To use \texttt{SecureInput} in other applications, such as password strength meter, the developers can customize the package to add the desired functionality. We note here that some features of password managers are not supported by \texttt{SecureInput}. For example, automatically reading and saving passwords from the login forms is not supported. We accept this as a limitation. However, this limitation can be addressed by asking the user to re-type the password in the password managers' input fields. The autofill feature of password managers remains unaffected.





\subsection{Built In}
\label{subsec:built-in}

The solution proposed above, \texttt{SecureInput}, acts as an add-on solution to prevent unrestricted access of sensitive input fields. However, this does not address the root cause of the vulnerability, i.e. lack of a fine-grained permission model for sensitive fields. Prior works~\cite{1_Liu2012ChromeET, 7_eriksson_hardening_2022} have proposed modifying the browser architecture to address this vulnerability. Specifically, Liu et al.~\cite{1_Liu2012ChromeET} suggest augmenting the existing privilege set to restrict Chrome extensions from accessing input fields of high `sensitivity.' Concurrently, Eriksson et al.~\cite{7_eriksson_hardening_2022} propose including a new \texttt{captureVisibleTab} permission to prevent extensions from obtaining user passwords. 

In this work, we take a different route and propose instrumenting Chrome to alert users whenever any JavaScript function accesses any password fields. We note here that instrumenting chrome is a big undertaking, and hence is out of scope for this work. Here, we present a proof-of-concept solution showcasing the necessary steps required to achieve the desired functionality. 

Our key insight here is that to programmatically access the sensitive values, the adversary first needs to select the element. We aim to intercept this access flow, and alert users when the access originates from JavaScript or browser extension. To develop a proof-of-concept solution, we focus on the flow where \texttt{document.querySelector} is used. We notify users both when the sensitive input field is selected and when its value is read. To accomplish this, we update the compiler file responsible for managing the document object to log a message whenever a querySelector selects the sensitive element. Finally, we update the core compiler for HTMLInputElement to log when the value of the sensitive field is read. \Cref{fig:app_built} shows the the logggin functionality on \url{facebook.com}. 

It's important to note that this represents only a proof-of-concept for a possible system that could be used to notify users. Updating Chromium to notify users with a more user-friendly design exceeds the scope of this paper. Furthermore, we only show the logging by intercepting one selection method \texttt{document.querySelector}, but the methodology can be extended to other selection methods as well. 

\noindent
\textbf{Impact on Type-A Vulnerability.} Recall that in \textit{Type-A} vulnerability, the password values are present in plain text in the source code of the element. Both the defenses presented in this section do not address this vulnerability as, if the password values are available in plain text, the adversary does not need to select the input element and read its value; they can simply copy the entire HTML and analyze the HTML on a remote server to extract the sensitive information.





\subsection{Tradeoffs}
\label{subsec:tradeoffs}

The bolt-on solution comprises a JavaScript library that keeps the password variable private, preventing JavaScript from accessing password values. This solution, which can be added to existing systems without major alterations to the underlying architecture, is fairly straightforward. It offers protection against numerous attacks that exploit JavaScript's access to password fields. However, the solution has its shortcomings. It doesn't guard against attacks that tamper with the entire HTML element or place a decoy textbox over the actual one, thereby leaving potential opportunities for exploitation. For example, the keylogging attack will bypass this defense and leak user data.

On the other hand, the built-in solution proposes a change at the browser's OS level and alerts users whenever an extension or JavaScript tries to access a sensitive field. This solution provides a more all-encompassing defense, tackling various potential attacks. Since it operates at the OS level, it offers a more cohesive and constant layer of protection. This solution can also integrate the functionality of password managers by giving trusted extensions exclusive access to password fields, thereby preserving usability while ensuring security. However, there are several challenges associated with this solution. Implementing the solution is more intricate and resource-consuming as an OS-level change than incorporating a bolt-on solution. It might also necessitate collaboration from various stakeholders, including browser developers, which could make the implementation process more complex.
\section{Related Works}
\label{sec:related-works}
\noindent
\textbf{Attacks involving Extension:} Prior studies~\cite{20_Wang_firefox_2012, 6_obimbo_analysis_2018, 7_eriksson_hardening_2022,24_louw_enhancing_2008} have detailed various techniques by which malicious extensions could leak sensitive information. Wang et al.~\cite{20_Wang_firefox_2012} conducted an empirical study on over 2400 Firefox extensions, revealing numerous vulnerabilities threatening web sessions. Obimbo et al.~\cite{6_obimbo_analysis_2018} showed how external attackers could take advantage of the elevated privileges given to extensions by exploiting code vulnerabilities present in those extensions. Similarly, Carlini et al.~\cite{9_carlini_evaluation_nodate} highlighted that due to the reality that not all extensions are developed by developers who are security experts, code vulnerabilities exist that can be exploited by malicious actors to expose user data potentially. Eriksson et al. \cite{7_eriksson_hardening_2022} systematically identified multiple attack entry points that a malicious browser extension could use to steal user information. For instance, they created a malicious extension that would alter the text type of password fields to `text’ and then capture a screenshot. Varshney et al.~\cite{26_Varshney_detecting_2017} demonstrated a variety of attacks a malicious extension could conduct to steal user information. These attacks included keylogging, credential sniffing, phishing, and inbox spying. However, all these attacks were not submitted as extensions to the webstore, making them theoretical rather than practically feasible, especially considering the current static and dynamic analysis defenses against such malicious extensions.

Bauer et al.~\cite{8_bauer_analyzing_2014} developed an iframe-based attack to stealthily steal user credentials by leveraging the autofill functionality of password managers. On the same lines, Perrorra et al.~\cite{4_perrotta_botnet_2018} crafted an extension that performed an iframe-based phishing attack where their extensions would fetch dynamic codes from a server and execute them. They managed to bypass and publish their extension to the Chrome web store. However, as we discuss later in \Cref{subsec:existing-threats}, this attack is no longer viable due to Chrome's ban on dynamic remote code execution.

In this work, we propose a new hybrid attack that leverages both static and dynamic attack techniques to extract sensitive user information. We further submit the extension to the Chrome web store and find that it bypasses the security checks, showing the practicality of the attack (see \Cref{sec:attack-design}).\smallskip

\noindent
\textbf{Detection of Malicious Extensions:} Several previous studies have devised tools and frameworks for the detection of malicious extensions. Research conducted in~\cite{11_zhao_automatic_2015, 13_wang_combined_2018} combined both static and dynamic analyses to identify and flag extensions. While Zhao et al.~\cite{11_zhao_automatic_2015} focused on the detection of information leaks via extensions, Wang et al.~\cite{13_wang_combined_2018} emphasized tracking DOM changes to identify malicious extensions. Varshney et al.~\cite{26_Varshney_detecting_2017} also introduced a static analysis framework for detecting malicious code within an extension. DeKoven et al.~\cite{12_dekoven_malicious_nodate} identified malicious extensions by flagging users who behave suspiciously on websites, subsequently scanning all loaded extensions for specific threat indicators. Shahriar et al.~\cite{10_shahriar_effective_2014} utilized a Hidden Markov Model to analyze and detect vulnerable and malicious extensions. Toreini et al.~\cite{21_Toreini_domtegrity_2019} created DOMtegrity to monitor and flag malicious DOM changes like `document.write’ or swapping child nodes. Previous research proposed dynamic analysis frameworks that analyze runtime code bases of extensions and match them to set heuristics to flag them as malicious~\cite{3_kapravelos_hulk_nodate, 5_chen_mystique_2018}.

Our work complements this line of study, offering a security analysis of vulnerabilities affecting browsers and generic solutions to address these vulnerabilities.

\noindent
\textbf{Browser Architecture:} Prior research has investigated how modifications to browsers' underlying structure can enhance user privacy and security. Louw et al.~\cite{24_louw_enhancing_2008} suggested incorporating a new runtime monitoring framework to observe an extension's access to sensitive APIs, such as adding an event listener to secure fields like passwords. Guha et al.~\cite{23_Guha_verified_2011}, and Liu et al.~\cite{1_Liu2012ChromeET} recommended adding new permissions to access specific DOM elements. Bauer et al.~\cite{8_bauer_analyzing_2014} explored how extensions could bypass the existing Chrome permission structure to execute a range of attacks. 
Some, though not all, of the changes advocated by these studies have been implemented in the latest iteration of the Chrome extension platform (Manifest V3), such as banning \texttt{eval} statements.

Unlike the aforementioned research that calls for a complete overhaul of the extension permission systems, we propose both a JS-based bolt-on solution and a Chromium patch that alerts users of sensitive field access (see \Cref{sec:remedies}). The chromium patch, distinct from the solution proposed by~\cite{1_Liu2012ChromeET, 23_Guha_verified_2011, 24_louw_enhancing_2008, 8_bauer_analyzing_2014, 9_carlini_evaluation_nodate}, does not necessitate any redesign of existing frameworks. Furthermore, the patch addresses threats from any malicious JS script and is not restricted solely to extensions.\smallskip


\section{Conclusion}

In this paper, we have presented a comprehensive analysis of the vulnerabilities associated with text input fields in web browsers, focusing on the exposure of sensitive information such as passwords, Social Security Numbers, and credit card details. We find that the lack of security boundary between the browser extension and the webpage results in novel vulnerabilities. Our case studies and large-scale measurements highlight the extent of these vulnerabilities, with alarming findings such as the exposure of passwords in plain text on over a 1000 websites, including popular ones like Google and Cloudflare. We also demonstrate the feasibility of a malicious extension bypassing existing protections and accessing sensitive data, underscoring the urgent need for more robust security measures.

To address these vulnerabilities, we propose two solutions: a JavaScript library that makes password variables private and a modified version of Chrome that notifies users when a password field is being accessed. While these solutions address some of the issues, they also highlight the need for a more comprehensive approach to securing sensitive input fields. 

\newpage
\bibliographystyle{plain}
\small{\bibliography{sample-base.bib}}

\begin{thebibliography}{10}

\bibitem{arsTechnicaAdwareMalware}
Ron Amadeo.
\newblock Adware vendors buy chrome extensions to send ad- and malware-filled
  updates.
\newblock
  \url{https://arstechnica.com/information-technology/2014/01/malware-vendors-buy-chrome-extensions-to-send-adware-filled-updates/},
  2014.

\bibitem{8_bauer_analyzing_2014}
Lujo Bauer, Shaoying Cai, Limin Jia, Timothy Passaro, and Yuan Tian.
\newblock Analyzing the dangers posed by chrome extensions.
\newblock In {\em 2014 IEEE Conference on Communications and Network Security},
  pages 184--192, 2014.

\bibitem{9_carlini_evaluation_nodate}
Nicholas Carlini, Adrienne~Porter Felt, and David Wagner.
\newblock An evaluation of the google chrome extension security architecture.
\newblock pages 97--111, August 2012.

\bibitem{5_chen_mystique_2018}
Quan Chen and Alexandros Kapravelos.
\newblock Mystique: Uncovering information leakage from browser extensions.
\newblock In {\em Proceedings of the 2018 ACM SIGSAC Conference on Computer and
  Communications Security}, CCS '18, page 1687–1700, New York, NY, USA, 2018.
  Association for Computing Machinery.

\bibitem{12_dekoven_malicious_nodate}
Louis~F. DeKoven, Stefan Savage, Geoffrey~M. Voelker, and Nektarios Leontiadis.
\newblock Malicious browser extensions at scale: Bridging the observability gap
  between web site and browser.
\newblock August 2017.

\bibitem{7_eriksson_hardening_2022}
Benjamin Eriksson, Pablo Picazo-Sanchez, and Andrei Sabelfeld.
\newblock Hardening the security analysis of browser extensions.
\newblock In {\em Proceedings of the 37th ACM/SIGAPP Symposium on Applied
  Computing}, SAC '22, page 1694–1703, New York, NY, USA, 2022. Association
  for Computing Machinery.

\bibitem{google_mv3_overview}
Google.
\newblock Overview of manifest v3.
\newblock
  \url{https://developer.chrome.com/docs/extensions/mv3/intro/mv3-overview/},
  2023.

\bibitem{23_Guha_verified_2011}
Arjun Guha, Matt Fredrikson, Benjamin Livshits, and Nikhil Swamy.
\newblock Verified security for browser extensions.
\newblock {\em 2011 IEEE Symposium on Security and Privacy}, pages 115--130,
  2011.

\bibitem{harrell_victims_2018}
Erika Harrell.
\newblock Victims of identity theft, 2018.
\newblock 2018.

\bibitem{Herley2012ARA_passwords_2}
Cormac Herley and Paul~C. van Oorschot.
\newblock A research agenda acknowledging the persistence of passwords.
\newblock {\em IEEE Security \& Privacy}, 10:28--36, 2012.

\bibitem{redditAdwareMalware}
Honey.
\newblock I am one of the developers of a popular chrome extension and we've
  been approached by malware companies that have tried to buy us. ama!
\newblock
  \url{https://www.reddit.com/r/IAmA/comments/1vjj51/i_am_one_of_the_developers_of_a_popular_chrome},
  2014.

\bibitem{Jagpal_2015}
Nav Jagpal, Eric Dingle, Jean-Philippe Gravel, Panayiotis Mavrommatis, Niels
  Provos, Moheeb~Abu Rajab, and Kurt Thomas.
\newblock Trends and lessons from three years fighting malicious extensions.
\newblock In {\em 24th USENIX Security Symposium (USENIX Security 15)}, pages
  579--593, Washington, D.C., August 2015. USENIX Association.

\bibitem{3_kapravelos_hulk_nodate}
Alexandros Kapravelos, Chris Grier, Neha Chachra, Christopher Kruegel, Giovanni
  Vigna, and Vern Paxson.
\newblock Hulk: Eliciting malicious behavior in browser extensions.
\newblock pages 641--654, August 2014.

\bibitem{cookieenforcer}
Rishabh Khandelwal, Asmit Nayak, Hamza Harkous, and Kassem Fawaz.
\newblock Cookieenforcer: Automated cookie notice analysis and enforcement.
\newblock {\em ArXiv}, abs/2204.04221, 2022.

\bibitem{lashkari2009shoulder}
Arash~Habibi Lashkari, Samaneh Farmand, Dr~Omar~Bin Zakaria, and Dr~Rosli
  Saleh.
\newblock Shoulder surfing attack in graphical password authentication.
\newblock {\em arXiv preprint arXiv:0912.0951}, 2009.

\bibitem{1_Liu2012ChromeET}
Lei Liu, Xinwen Zhang, Guanhua Yan, and Songqing Chen.
\newblock Chrome extensions: Threat analysis and countermeasures.
\newblock In {\em Network and Distributed System Security Symposium}, 2012.

\bibitem{24_louw_enhancing_2008}
Mike~Ter Louw, Jin~Soon Lim, and Venkat Venkatakrishnan.
\newblock Enhancing web browser security against malware extensions.
\newblock {\em Journal in Computer Virology}, 4:179--195, 2008.

\bibitem{6_obimbo_analysis_2018}
Charlie Obimbo, Yong Zhou, and Randy Nguyen.
\newblock Analysis of vulnerabilities of web browser extensions.
\newblock In {\em 2018 International Conference on Computational Science and
  Computational Intelligence (CSCI)}, pages 116--119, 2018.

\bibitem{4_perrotta_botnet_2018}
Raffaello Perrotta and Feng Hao.
\newblock Botnet in the browser: Understanding threats caused by malicious
  browser extensions.
\newblock {\em IEEE Security \& Privacy}, 16(4):66--81, 2018.

\bibitem{27_Richards_eval_2011}
Gregor Richards, Christian Hammer, Brian Burg, and Jan Vitek.
\newblock The eval that men do: A large-scale study of the use of eval in
  javascript applications.
\newblock In {\em Proceedings of the 25th European Conference on
  Object-Oriented Programming}, ECOOP'11, page 52–78, Berlin, Heidelberg,
  2011. Springer-Verlag.

\bibitem{10_shahriar_effective_2014}
Hossain Shahriar, Komminist Weldemariam, Mohammad Zulkernine, and Thibaud
  Lutellier.
\newblock Effective detection of vulnerable and malicious browser extensions.
\newblock {\em Computers \& Security}, 47:66--84, 11 2014.

\bibitem{21_Toreini_domtegrity_2019}
Ehsan Toreini, Maryam Mehrnezhad, Siamak~Fayyaz Shahandashti, and Feng Hao.
\newblock Domtegrity: ensuring web page integrity against malicious browser
  extensions.
\newblock {\em International Journal of Information Security}, 18:801 -- 814,
  2019.

\bibitem{22_Ursell_desktop_2019}
Steven Ursell and Thaier Hayajneh.
\newblock Desktop browser extension security and privacy issues.
\newblock {\em Lecture Notes in Networks and Systems}, 2019.

\bibitem{26_Varshney_detecting_2017}
Gaurav Varshney, Manoj Misra, and Pradeep Atrey.
\newblock Detecting spying and fraud browser extensions: Short paper.
\newblock pages 45--52, 10 2017.

\bibitem{20_Wang_firefox_2012}
Jiangang Wang, Xiaohong Li, Xuhui Liu, Xinshu Dong, Junjie Wang, Zhenkai Liang,
  and Zhiyong Feng.
\newblock An empirical study of dangerous behaviors in firefox extensions.
\newblock pages 188--203, 09 2012.

\bibitem{13_wang_combined_2018}
Yao Wang, Wandong Cai, Pin Lyu, and Wei Shao.
\newblock A combined static and dynamic analysis approach to detect malicious
  browser extensions.
\newblock {\em Security and Communication Networks}, 2018:1--16, 05 2018.

\bibitem{Zhao2013VulnerabilityAR_Passwords_1}
Rui Zhao, Chuan Yue, and Kun Sun.
\newblock Vulnerability and risk analysis of two commercial browser and cloud
  based password managers.
\newblock {\em Science}, 2:183--197, 2013.

\bibitem{11_zhao_automatic_2015}
Rui Zhao, Chuan Yue, and Qing Yi.
\newblock Automatic detection of information leakage vulnerabilities in browser
  extensions.
\newblock In {\em Proceedings of the 24th International Conference on World
  Wide Web}, WWW '15, page 1384–1394, Republic and Canton of Geneva, CHE,
  2015. International World Wide Web Conferences Steering Committee.

\end{thebibliography}

\newpage
\appendix

\newpage
\section{Appendix}
\label{sec:appendix}



\subsection{Chrome Instrumentation}
\label{subsec:chrome_instrumentation}

\begin{figure*}[h]
  \centering
  \includegraphics[width=\textwidth]{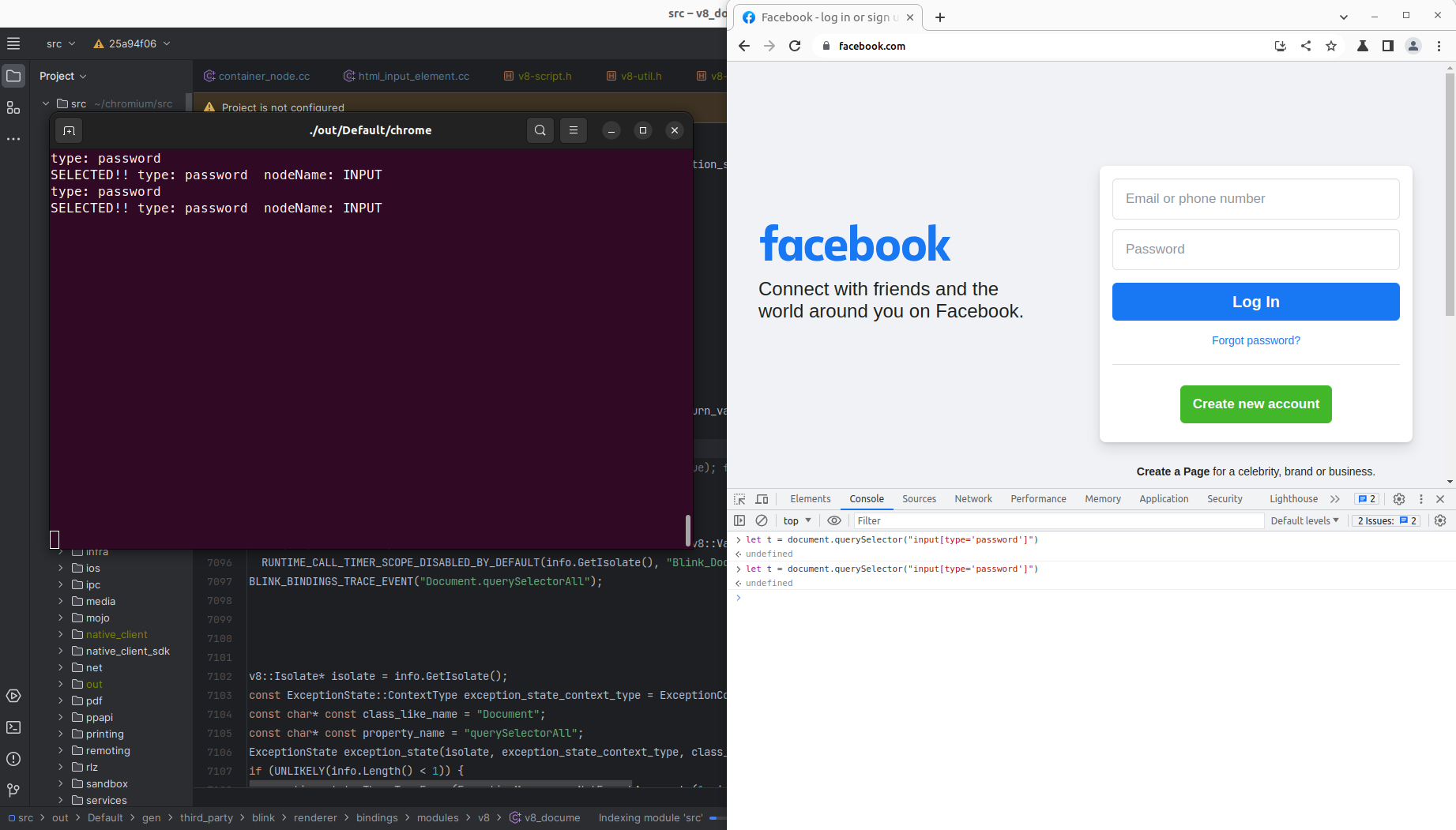}
  \caption{The output of the logging code as a part of our chrome instrumentation to intercept sensitive element selection and notify users.}
  \label{fig:app_built}
\end{figure*}

\end{document}